\documentclass[twocolumn,prb,showpacs,amsmath,amssymb]{revtex4}
\usepackage{graphicx}% Include figure files
\usepackage{dcolumn}% Align table columns on decimal point
\usepackage{bm}% bold math
\begin{document}
\title{Bosonic versus fermionic pairs of topological
spin defects in monolayered high-\(T_\text{c}\) superconductors}
%\shorttitle{Bosonic vs fermionic pairs ...}

\author{M.A. Garcia-Bach}%
\affiliation
{Departament de F\'{\i}sica Fonamental, Facultat de
F\'{\i}sica, Universitat de Barcelona, Diagonal 647,
E-08028 Barcelona, Catalunya, Spain.   \\
\email:  {m.angels.garcia\_bach@ub.edu}}

\date{\today}

\begin{abstract}
The energy associated with bosonic and fermionic pairs
of topological spin defects in doped antiferromagnetic quantum
spin-1/2 square lattice is estimated within a resonating
valence bond scenario, as described by a $t$-$t'$-$J$-like
model Hamiltonian, plus a \(t_\perp\),
responsible of a three-dimensional screening of the
electrostatic repulsion within the bosonic pairs.
For parameters appropriate for monolayered high-\(T_\text{c}\)
superconductors, both fermionic and bosonic pairs show $x^2-y^2$
symmetry.
We find a critical value of doping such that the energy of the bosonic
pairs goes below twice the energy of two fermionic pairs
at their Fermi level.
This finding could be related to the onset of high-$T_\text{c}$
superconductivity.
\end{abstract}


\maketitle

\section{Introduction}
\label{sec:intro}

Since the discovery of high-$T_\text{c}$
superconductivity~(HTSC) in La$_2$CuO$_4$,\cite{Bednorz} a vast
amount of work has been done on slightly doped quasi-two
dimensional~(2D)
antiferromagnets.\cite{Barnes,Dagotto94,Orenstein,NCYeh,Norman}
It has been observed that these materials display very unusual
properties, with a rich variety of temperature-doping phases diagram.
Specifically, away from the overdoped side, the cuprates do not
appear to be a Landau Fermi liquid.
For instance, they should be considered as doped Mott insulators.
However, the theoretical status of the field has been largely
phenomenological and controversial.\cite{NCYeh,Norman}
As far as we know, there is no consensus on the origin of the
superconductivity nor on the pseudogap phase.
Therefore, finding a microscopic mechanism for HTSC still
is an open problem.

Local-density approximation (LDA) and generalized gradient
approximation (GGA)\cite{Perdew} to density functional theory have
been used so far to rationalize the electronic structure of HTSCs.
Although LDA is a useful technique for some materials, it has been
shown that both LDA or GGA are not appropriate for
antiferromagnetic materials because they tend to yield a metallic
ground state with incorrect delocalized spin density and band
ordering.\cite{Pickett,Moreira}
This is attributed to an extreme nonanalytic
and nonlocal behavior of density functional theory as the particle
number is changed,\cite{Godby,Hybertsen86} implying the need for a
self-energy correction, or at least an orbital dependent potential to
obtain a realistic description of band gaps.
To overcome such a problem, different semiempirical corrections
to LDA  have been proposed so far as, i.e.,
LDA+SIC (Refs.~\onlinecite{S-Gunnarsson,Svane,Szotek} and
LDA+$U$.\cite{Czyzyk,Anisimov,Wei}

An alternative approach to the electronic structure of the
HTSCs is based on the use of model Hamiltonians that aim to
incorporate the essential physics into a few parameters.
It is generally accepted that electron correlation is important for
HTSC.
Furthermore, it is well known that the (covalent-structure)
valence-bond~(VB) model or, equivalently, the Heisenberg Hamiltonian
includes most of the electron correlation. 
Thence, early in 1987, Anderson\cite{Anderson87} proposed that
the important features of the undoped HTSC parent compounds can
be described by a Heisenberg Hamiltonian on a two-dimen\-sional
square lattice with one electron per site.
Meanwhile, Emery\cite{Emery87,Emery88} proposed a three-band Hubbard
model.
Unfortunately, the number of parameters of a three-band Hubbard model
turns to be too large.
Therefore, Zhang and Rice\cite{Zhang} proposed a simplification of
the three-band Hubbard model into the well known $t$-$J$,
which implicitly includes the O($p$)-Cu($d$) hybridization and recovers
the initial effective one-band description of Anderson.
Since an appropriate parametrization is essential for the predictive
capability of model Hamiltonians, much progress has been achieved
on the high-level
\emph{ab initio} computation of reliable appropriate
parameters using only the crystal structure as external
input.\cite{CPL,Calzado,Munoz}

Stimulated by Anderson's suggestion,\cite{Anderson87} a renewed
interest of low-dimensional quantum spin-1/2 antiferromagnetic
systems emerged.
According to the Lieb and Mattis theorem\cite{Lieb-Mattis}
the ground state for the undoped half-filled bipartite system
must be a singlet.
Therefore, the appropriate ground-state wave function could have a
resonat\-ing-valence-bond~(RVB) character.
It was soon pointed out that short-range RVB wave functions exhibit
topological long-range order.\cite{KivelsonRS,Kivelson,PRB-91,KleinZV}
Furthermore, recently\cite{Hansson} topological order for
superconductors has also been claimed, away from truly microscopic
models, making use of bosonic theories of the quantum Ginzburg-Landau
form.
In Ref.~\onlinecite{PRB-91} Klein and collaborators investigated
the short-range RVB wave functions within a dimer coverings
approximation for the square lattice and found that the
dimer-coverings show a type of long-range spin-pairing order (LRSPO).
Using arguments based on the LRSPO they predicted a per-site
energy $\varepsilon \propto \delta^2$, where $\delta$ is
the deviation of the local LRSPO with respect to the LRSPO of the
ground state.
Furthermore, topological spin defects~(TSDs), namely a site that
is not spin paired to a singlet, or a hole in hole-doped
superconductors, or a doubly occupied site in electron-doped materials,
were assimilated to Bloch walls separating phases with a difference in
LRSPO of $\pm 1$.
It was argued that, at a longitudinal distance $\sim \rho$ past the
TSD on the less stabilized side, the defect should also presumably have
only spread out a transverse distance $\sim \rho$, so that
$\delta \sim 1/\rho$, and $\varepsilon \propto 1/\rho^2$.
Therefore, the energy contribution from all the sites of the given
longitudinal distance past the TSD is
$\rho \Delta\varepsilon \sim 1/\rho$.
Summation over all the sites up to a given distance thence gives an
energy cost of $\sim\ln \rho$.
When the TSD are charged, it was also suggested that this long flat
attraction $\sim\ln \rho$ along with the screened repulsion
$\exp \{-\alpha \rho \} /\rho$ could lead to a weakly bound pair.
Recently,\cite{CPL} a linear relationship between $T_c$
and the $J/t$ ratio, as obtained from high-level
\emph{ab initio} calculations, was found.
It was argued that such a linear relation arises from the LRSPO
mechanism previously suggested.\cite{PRB-91}
Furthermore, the so-called $t$-$J$ Hamiltonian for the cuprates seems
to be pointing to the right direction.

The existence of a LRSPO for more general RVB wave functions has
been proven for ladderlike quantum spin-1/2 antiferromagnetic
systems.\cite{EPJ,capitolVB}
Most of the considerations associated to the existence of this LRSPO
for the ladderlike quantum spin-1/2 antiferromagnetic systems are
readily applicable to the square lattice.
In particular, bound pairs of TSD are predicted to occur.
However, as far as we know, the energy of such a pair of TSD
as a function of the distance has not been obtained yet.
Even more, arguments based on the LRSPO alone cannot decide if
vacancies (doubly occupied sites), let's say
\emph{charge-wearing} TSDs, organize themselves as bound pairs of two
charge-wearing TSDs, as bosonic-character pairs, or each
charge-wearing TSD would bind to a non spin-paired spin, let us say a
\emph{spin-wearing} TSD, leading to a fermionic-character pair.

Here focus is on the energy associated with these
bosonic and fermionic pairs as described by symmetry-adapted
extended wave functions.
We find that the fermionic pairs are favored for low doping levels,
but the Fermi level increases with doping while the energy of the
bosonic pairs lowers.
At a critical doping the energy of the bosonic pair goes below the
energy of two fermionic pairs at the Fermi level, suggesting the
pairing of charge-wearing TSDs, and hence providing a microscopic
mechanism for HTSC.

The description of these bosonic and fermionic pairs is based on a
$t$-$t'$-$J$-like model Hamiltonian $H = H_I + H_J + H_t + H_{t'}$,
where $H_I$ is the energy associated with the ionization potential for
hole-doped materials, or the energy associated with the electron
affinity for the electron-doped systems.
The $H_J$ is the well known nearest-neighbor Heisenberg Hamiltonian,
\begin{equation}
H_J = J \sum_{\langle \mathbf{R},\mathbf{R}'\rangle }
\mathbf{S}_{\mathbf{R}} \cdot \mathbf{S}_{\mathbf{R}'},
\end{equation}
where $\mathbf{S}_{\mathbf{R}}$ is the spin operator for the spin on
the site $\mathbf{R}$, and $\langle \mathbf{R},\mathbf{R}'\rangle$
means that $\mathbf{R}$ and $\mathbf{R}'$ are nearest neighbors.
The nearest- and next-nearest-neighbor hopping contributions to the
Hamiltonian are, respectively,
\begin{eqnarray}
H_t &=& -t \sum_{\langle \mathbf{R},\mathbf{R}'\rangle } \sum_\sigma
\left( c^\dagger_{\mathbf{R}\sigma} c_{\mathbf{R}'\sigma}
+ c^\dagger_{\mathbf{R}'\sigma} c_{\mathbf{R}\sigma} \right)
\nonumber \\[2mm]
&\times & \left( 1-\hat{n}_{\mathbf{R}\bar{\sigma}}\right)
\left( 1-\hat{n}_{\mathbf{R}'\bar{\sigma}}\right) , \\
H_{t'} &=& \sum_{\langle\langle \mathbf{R},\mathbf{R}'\rangle\rangle }
t'_{\langle\langle \mathbf{R},\mathbf{R}'\rangle\rangle }
\sum_\sigma \left( c^\dagger_{\mathbf{R}\sigma} c_{\mathbf{R}'\sigma}
+ c^\dagger_{\mathbf{R}'\sigma} c_{\mathbf{R}\sigma} \right)
\nonumber \\[2mm]
&\times & \left( 1-\hat{n}_{\mathbf{R}\bar{\sigma}}\right)
\left( 1-\hat{n}_{\mathbf{R}'\bar{\sigma}}\right) ,
\end{eqnarray}
where $c^\dagger_{\mathbf{R}\sigma}$ ($c_{\mathbf{R}\sigma}$) creates
(destroys) an electron on site $\mathbf{R}$ with spin
$\sigma = \alpha$, $\beta$.
The double occupancy is avoided by the factors
$1-\hat{n}_{\mathbf{R}\bar{\sigma}}$, where
$\hat{n}_{\mathbf{R}\bar{\sigma}}$ is the number operator on site
$\mathbf{R}$ with spin $\bar{\sigma} = \beta$, $\alpha$. 
The summation on $\langle\langle \mathbf{R},\mathbf{R}'\rangle\rangle$
means that $\mathbf{R}$ and $\mathbf{R}'$ are next-nearest neighbors.
The hopping integral
$t'_{\langle\langle \mathbf{R},\mathbf{R}'\rangle\rangle }$ depends on
the number of holes within the plaquette. 
Such a model is known to reproduce the low-energy
spectrum of the three-band Hubbard model.\cite{Hybertsen90}
Here, we use the parameters obtained by high-level
\emph{ab initio} calculations using only the crystal structure as
external input.\cite{CPL,Calzado}
We approximate the screened electrostatic repulsion within the bosonic
pair by the Yukawa potential,\cite{Ashcroft,Madelung}
the screening agent being the gas of the fermionic pairs.
The three-dimensional (3D] character of the screening is taken into
account by an interlayer hopping integral $t_\perp$.
The Heisenberg part of the energy associated with a pair of static
TSDs is estimated by the dimer-covering-\emph{counting}
approximation\cite{Seitz,Klein-86,Zivkovic}
on $w\times L$ antiferromagnetic quantum spin-1/2 square lattice, with
$w= 4, 6, \dots , 20$, $L\rightarrow\infty$, and cyclic boundary
conditions in both directions.
Counting of the dimer-covering configurations has been achieved by
a transfer-matrix technique.\cite{EPJ,capitolVB}

This paper is organized as follows:
In Sec.~\ref{sec:SPO} we review the main concepts about
LRSPO,\cite{EPJ,capitolVB} describing the scenario where the TSDs
are located.
In Sec.~\ref{sec:counting} the energy per site of the half-filled
ground state, and the gain of the Heisenberg energy associated with a
static pair of TSDs is estimated.
In Sec.~\ref{sec:2D-wf} the symmetry-adapted extended wave-functions
appropriate for bosonic and fermionic pairs moving in a CuO$_2$ layer
will be defined, and the energy bands will be obtained.
From the two-fluids equilibrium condition, the critical doping $p_c$
for the onset of pairing to bosonic pairs among charge-wearing TSDs is
obtained.
Finally a summary and the conclusions are given in
Sec.~\ref{sec:conclusions}.

\section{Long-range spin-pairing order
and topological spin defects}
\label{sec:SPO}

\begin{center}
\begin{figure}
\includegraphics[width=8.5cm]{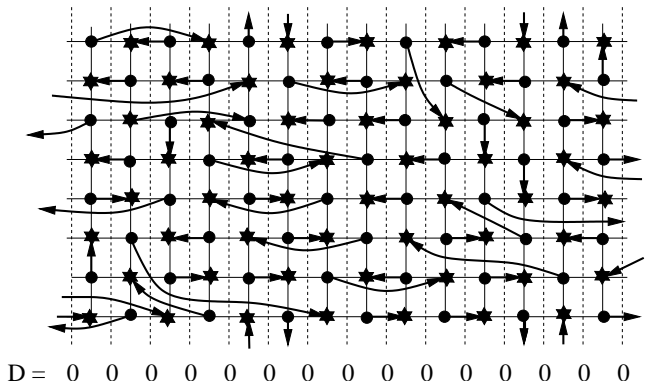}
\caption{A fragment of a VB configuration for a $w=8$ square lattice.
Each arrow represents a spin-pairing (SP) to a singlet between a spin
on a site in the sublattice $\mathcal{A}$ (circles) with a spin on a
site in the sublattice $\mathcal{B}$ (stars).
Below each boundary, i.e.\ the dashed lines running parallel to the
rungs, it appears the net count of arrows, $D$, penetrating this
boundary.}
\label{fig:LRSPO}
\end{figure}
\end{center}

From the Lieb and Mattis theorem\cite{Lieb-Mattis} it is
well known that for bipartite spin systems a
\emph{maximally-spin-paired} ground state is expected.
In particular, at half filling, for ladder-like systems, with equal
number of sites in the $\mathcal{A}$ and $\mathcal{B}$ sublattices,
the ground state is a singlet.
Singlet states can be achieved by configuration interaction (CI)
among covalent VB configurations or RVB.
For instance, a linearly independent set of VB configurations can be
achieved by pairing to a singlet each spin in the sublattice
$\mathcal{A}$ to a spin in the sublattice $\mathcal{B}$, not
necessarily one of its nearest neighbors (see Fig.~\ref{fig:LRSPO}).

It is known\cite{EPJ,capitolVB} that any (covalent) VB configuration
exhibits a LRSPO related to the local (at boundary) array of SPs
penetrating any boundary $f_n$ (see, for instance,
Fig.~\ref{fig:LRSPO}).
The parameter associated with the LRSPO, $D$, can take $w+1$ different
relevant values.
The shape of the boundary selected to define the different $w+1$
values of $D$ is quite arbitrary, though when $w$=even and the
boundary is chosen to run parallel to rungs, the $w+1$~different
values of $D$ are:
\begin{equation}
D = 0, \pm 1, \dots , \pm \frac{w}{2}.
\end{equation}
This LRSPO allows to separate the set of VB configurations in
different subsets.
Since two singlets from different subsets must be different repeatedly
at every position along the ladder, they are asymptotically
orthogonal and non interacting via any interaction mediated by a
few-particle operator.
Then the matrix of the Hamiltonian asymptotically block-diagonalizes,
so configurations belonging to different subsets do not mix in the CI
sense.
Thus $D$ may be taken as a long-range order parameter labelling
the eigenstates of the $D$ block.
Under low-frustration conditions, the expected ordering of the
lowest-lying energy $E_D$ from the different blocks is
\begin{equation}
E_0 < E_1 < \cdots < E_{w/2},
\label{eqn:ordenacio-2}
\end{equation}
with $E_D = E_{-D}$.

Now, half-filled excited states or slightly doped states are
analyzed via TSDs.
There are different types of excitations conceivable from a
\emph{maximally-spin-paired} ground state.  Let us say, preserving half
filling (one electron per site), there are primarily spin excitations.
In this case, two spin-wearing TSDs, one in the sublattice
$\mathcal{A}$ and the other in the sublattice $\mathcal{B}$, are
obtained by breaking one SP to form a triplet state.
Away from half-filling, there are low-energy spin and charge
excitations.
For instance, removing (adding) one electron produces two sites that
cannot be SP, a charge-wearing TSD and a spin-wearing TSD, one in the
sublattice $\mathcal{A}$ and the other in the sublattice $\mathcal{B}$,
the ladder becoming a doublet.
In this case hopping terms must be retained in the Hamiltonian and
the so-called \(t\)-\(J\) model or different
extensions that incorporate either next-nearest-neighbor hopping
$t'$ or electrostatic repulsion have been employed so far.
Thence, the doublet is a weighted superposition of VB configurations
with a spin-wearing TSD and a charge-wearing TSD lying in different
sublattices.
Still, going up in the hierarchy of Hamiltonians, the Hubbard or even
a more general Hamiltonian must be considered.
In this case, still another type of excitations (though presumably of
higher energy if a Heisenberg-type Hamiltonian is assumed to govern
the lowest-lying region of the spectrum) can be produced relaxing the
single-occupancy constrain.  This leads to the \emph{ionic} states,
i.e., states with at least a pair of sites, one doubly occupied
and the other empty, namely one negatively charge-wearing TSD and one
positively charge-wearing TSD.

\begin{center}
\begin{figure}
\includegraphics[width=8.5cm]{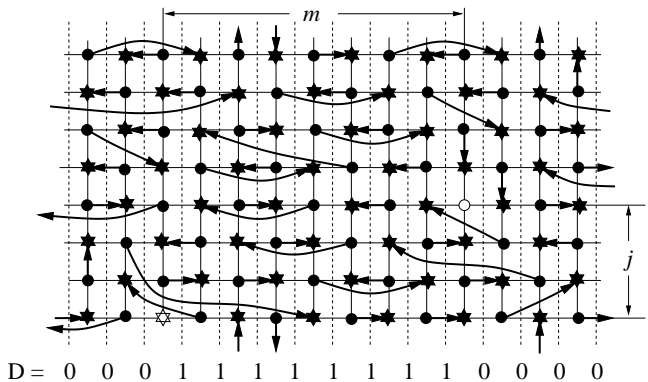}
\caption{A fragment of a VB configuration for a $w=8$ square lattice
containing a pair of TSDs, one in the sublattice $\mathcal{A}$ (white
circle), and the other in the sublattice $\mathcal{B}$ (white star).
Notice that this VB configuration shows LRSPO $D=0$
everywhere but in the intervening region defined by the pair of TSDs,
with $D=1$.}
\label{fig:LRSPO-parell}
\end{figure}
\end{center}

Of special relevance here is how the LRSPO is disrupted by a TSD (see
Fig.~\ref{fig:LRSPO-parell}).
For instance, a TSD in a site $[n,i]$, $n$ indicating the rung and $i$
the leg, can be seen as a domain wall
on the rung $n$ which separates the lattice in two sectors with
associated left, $D_l$, and right, $D_r$, order parameters.
When we choose the sublattice $\mathcal{A}$ as formed by the
set of sites $[m,j]$ with $m+j$=even,
\begin{equation}
D_r = D_l -(-1)^{n+i}.
\label{eqn:discontinuitat}
\end{equation}
Thence, to fulfill boundary conditions TSDs must appear by pairs, one
TSD in the sublattice $\mathcal{A}$ and the other in the sublattice
$\mathcal{B}$, to ensure $\Delta D = 0$ from the left to the right
of the pair.
Such a pair define an intervening region with $\Delta D = \pm 1$ with
respect to the LRSPO $D$ of the host (see
Fig.~\ref{fig:LRSPO-parell}).
Then, away from half-filling, it may be conceivable an intervening
region limited by two charge-wearing TSDs, or a charge-wearing TSD and
a spin-wearing TSD (provided that the doping is not so strong as to
preclude a maximally-spin-paired ground state).
In particular, when placing a pair of TSDs above the ground state
($D$=0), the order parameter of the intervening region will be
$|D_p|$=1, which from Eq.~(\ref{eqn:ordenacio-2}) is expected to have
higher associated energy per site.
This indicates that the pair of TSDs should try to remain as close as
possible.
Thus, bound pairs of TSDs are predicted to occur.
To show that this is the case is one of the concerns of the present
paper.

\section{Heisenberg energy of a static pair of TSDs}
\label{sec:counting}

Within the dimer-covering-\emph{counting} approximation the
\emph{resonance energy}, $E_r(w,D)$ in units of $J$,
i.e., the energy correction below the energy of a
single dimer-covering structure, depends on the
configuration interaction among the different dimer-covering
configurations with LRSPO $D$.
When an equally weighted wave function is considered,
it has been argued\cite{Seitz,Klein-86,Zivkovic} that one might
consider this energy lowering to depend solely on the dimension of
the space spanned by the appropriate dimer-covering configurations.
Let $\mathcal{N}_D(w)$ be the number of linearly independent
dimer-covering configurations with the LRSPO $D$.
Since $\mathcal{N}_D(w)$ is multiplicative in terms of a break up into
subsystems while the energy is additive, such a functional dependence
should be of the form
\begin{equation}
E_r(w,D) \approx -C \ln \mathcal{N}_D(w),
\label{eqn:E-r}
\end{equation}

The values $\mathcal{N}_D(w)$ can be easily obtained by a
transfer-matrix technique.\cite{EPJ}
Let us start computing $\mathcal{N}_D(w)$ for a maximally spin-paired
half-filled system.
Let us analyze from a local point of view the dimer-covering singlets.
We can identify the dimer-covering local states according to which legs
have a pairing across the $f_n$ boundary.
In the present case it can be seen that, for each boundary, there are
$2^w$ different local states, $| e_{nI} )$ ($I$ ranging), which can be
classified according to the value of $D$, $| e^D_{nI} )$.

Proceeding from the left to the right, from the boundary $f_{n-1}$ to
$f_n$, a dimer-covering-\emph{counting} matrix, ${\cal K}_n$, is
defined as $(e_{n-1 I}| {\cal K}_n | e_{n J}) \equiv $ the number of
different ways $| e_{n J})$ can succeed $| e_{n-1 I})$.
Then, the number of dimer-covering states in a $D$ subspace is
\begin{equation}
\mathcal{N}_D(w) = \sum_{e^D_{0I}}
( e^D_{0I}| {\cal K}_1{\cal K}_2 \cdots {\cal K}_L | e^D_{0I}).
\end{equation}
Since $D_n = D_{n+1}$ for any dimer-covering singlet,
${\cal K}_n$ is a block-diagonal symmetric matrix that
does not depend on $n$ and we can omit this subindex.  For
$L\rightarrow\infty$, the highest eigenvalue $\Lambda_{wD}$ of the
$D$ block ${\cal K}_D$ dominates, and
\begin{equation}
\mathcal{N}_D(w) \approx \Lambda^L_{wD}.
\end{equation}
Therefore,
\begin{equation}
E_D(w) \approx w L\varepsilon_0 +E_r(w,D)
\approx w L\varepsilon_0 -CL \ln \Lambda_{wD},
\label{eqn:energyD}
\end{equation}
where $\varepsilon_0=-0.375$ is the energy per spin of a single
dimer-covering configuration.
Since\cite{EPJ,capitolVB}
\begin{equation}
\Lambda_{wD} > \Lambda_{wD'} \quad \text{when} \quad |D| < |D'|,
\end{equation}
with $\Lambda_{wD} = \Lambda_{w |D|}$,
the Heisenberg energy for the half-filled ground state belongs to the
subspace $D$=0, as suggested by Eq.~(\ref{eqn:ordenacio-2}), and can be
approximated (in units of $J$) by
\begin{equation}
E_0(w) \approx w L\varepsilon_0 -CL \ln \Lambda_{w0}.
\label{eqn:energy0}
\end{equation}

$C$ is a fitting parameter independent of the structure to
some degree.
The value of $C$ for the nearest-neighbor isotropic
Heisenberg model has been determined for a class of benzenoid
hydrocarbons\cite{Seitz} (with $C$=0.5667) and for finite
square-lattice fragments\cite{Zivkovic} (with $C$=0.75), by
fitting the logarithm of the dimer-covering count to the
resonance energy calculated from an equally weighted
dimer-covering wave function.
For the spin-1/2 square lattice, more general RVB approximations
suggest\cite{EPJ} a rough estimate of $C = 0.94 \pm 0.19$.
Here $C$ is fixed to yield a reasonably good estimate of the
ground-state Heisenberg energy of the half-filled square lattice.
Table~\ref{tab:fonamental} summarizes the ground-state energy
per site for $w=4, 6, \dots , 20$ and its extrapolation to
$w \rightarrow \infty$.
We use $C = 1$ from here on, since $C \approx 1.0083$ yields the
ground-state energy of Liang \emph{et al.}\cite{Liang}
\begin{table}
\caption{The ground-state resonance energy per site in units of
$CJ$ $(-\ln \Lambda_{w0}/w$), and the extrapolation to
$w \rightarrow \infty$. In the third column the ground-state
energy per site in units of $J$ as obtained when taking
$C=1$ .\protect\label{tab:fonamental}}
%\begin{ruledtabular}
\begin{center}
\begin{tabular}{ccc}
$w$ & $-\left( \ln \Lambda_{w0}\right) /w$ & $\varepsilon_0$\\
\hline
4  & -0.3292  & -0.7042\\
6  & -0.3073  & -0.6823\\
8  & -0.3001  & -0.6751 \\
10 & -0.2969 & -0.6719 \\
12 & -0.2953 & -0.6703 \\
14 &-0.2943  & -0.6693 \\
16 & -0.2936 & -0.6686 \\
18 & -0.2932 & -0.6682\\
20 & -0.2929  &  -0.6679\\
$\infty$ &-0.2913  & -0.6664 \\
\end{tabular}
\end{center}
%\end{ruledtabular}
\end{table}

When adding a TSD to a CuO$_2$ layer the transfer matrix ${\cal K}$
across the defect must be substituted by the appropriate
${\cal K}_{\mathbf{R}}$, where $\mathbf{R}$ is the vector position of
the TSD.
Therefore, the number of dimer-covering configurations when adding a
pair of TSDs to the half-filled ground state, located, respectively,
at $\mathbf{0}$ and $[m,j]$, with non-negative $m$ and $j$, with
$m+j$=odd, is
\begin{eqnarray}
\mathcal{N}_{[m,j]}(w) &= \left(\Lambda_{w0}| 
{\cal K}_{\mathbf{0}}{\cal K}^{m-1}{\cal K}_{[m,j]}
{\cal K}^{L-m-1} |\Lambda_{w0}\right) \nonumber  \\
& \approx  \Lambda_{w0}^{L-m-1}
(\Lambda_{w0}| {\cal K}_{\mathbf{0}}{\cal K}^{m-1}{\cal K}_{[m,j]}
|\Lambda_{w0}).
\end{eqnarray}
Thence, the Heisenberg energy (in units of $J$) associated with a pair
of static TSDs separated $[m,j]$, $m+j$=odd, with respect to the
energy of the half-filled ground state is
\begin{equation}
\varepsilon_{[m,j]} (w) \approx  -2 \varepsilon_0
+ \ln \frac{\Lambda^{m+1}_{w0}}
{( \Lambda_{w0}| {\cal K}_{\mathbf{0}}{\cal K}^{m-1}{\cal K}_{[m,j]}| 
\Lambda_{w0})} .
\label{eqn:energy-mj}
\end{equation}

Table~\ref{tab:energy-mj} summarizes the energies
$\varepsilon_{[m,j]}(w)$ from $[m,j]=[1,0]$
to $[7,4]$ and $w=4$, 6, $\cdots$, 20.
The $w \rightarrow \infty$ limit, $\varepsilon_{[m,j]}$,
is obtained by fitting $\varepsilon_{[m,j]}(w)$ by a power
series in $1/w$.
\begin{table*}
\caption{Energy with respect to the half-filled ground state,
$\varepsilon_{[m,j]}(w)$, of Eq.~(\ref{eqn:energy-mj}),
associated with a pair of TSDs separated $[m,j]$, and the
extrapolation to $w \rightarrow \infty$, $\varepsilon_{[m,j]}$.
Values for $[0,1]$ and $[1,2]$ have been included to emphasize that
$\varepsilon_{[j,m]}$ is asymptotically equivalent to
$\varepsilon_{[m,j]}$.\protect\label{tab:energy-mj}}
%\begin{ruledtabular}
\begin{tabular}{ccccccccccc}
$[m,j]$& $w=$ 4 & 6 & 8 & 10 & 12 & 14 & 16 & 18 & 20 & $\infty $\\
\hline\noalign{\smallskip}
$[1,0]$ & 2.3044 & 2.2064 & 2.1730 & 2.1587 & 2.1515
        & 2.1473 & 2.1447 & 2.1429 & 2.1416 & 2.1334 \\
$[0,1]$ & 1.9925 & 2.0708 & 2.1009 & 2.1143 & 2.1213
        & 2.1254 & 2.1280 & 2.1298 & 2.1310 & 2.1386  \\
$[2,1]$ & 2.8537 & 2.6837 & 2.6325 & 2.6133 & 2.6044
        & 2.5996 & 2.5966 & 2.5947 & 2.5933 & 2.5792   \\
$[1,2]$ & 2.5418 & 2.5481 & 2.5605 & 2.5689 & 2.5742
             & 2.5776 & 2.5710 & 2.5816 & 2.5827 & 2.5845 \\
$[3,0]$ & 3.2688 & 2.9316 & 2.8071 & 2.7508 & 2.7210
        & 2.7034 & 2.6921 & 2.6845 & 2.6790 & 2.6509    \\
$[3,2]$ & 3.3094 & 3.0050 & 2.9020 & 2.8594 & 2.8385
        & 2.8268 & 2.8196 & 2.8148 & 2.8114 & 2.7900      \\
$[4,1]$ & 3.7245 & 3.2529 & 3.0766 & 2.9969 & 2.9551
        & 2.9307 & 2.9151 & 2.9045 & 2.8971 & 2.8617      \\
$[4,3]$ &             & 3.2657 & 3.1062 & 3.0416 & 3.0115
        & 2.9956 & 2.9863 & 2.9804 & 2.9764 & 2.9481       \\
$[5,0 ]$& 4.1566 & 3.5239 & 3.2728 & 3.1525 & 3.0863
        & 3.0461 & 3.0199 & 3.0018 & 2.9889 & 2.9304    \\
$[5,2]$ & 4.1636 & 3.5393 & 3.2961 & 3.1823 & 3.1210
        & 3.0845 & 3.0611 & 3.0452 & 3.0338 & 2.9866    \\
$[6,1]$ & 4.5957 & 3.8103 & 3.4923 & 3.3379 & 3.2522
        & 3.2000 & 3.1659 & 3.1424 & 3.1256 & 3.0553   \\
$[5,4]$ &             &             & 3.3104 & 3.2115 & 3.1629
        & 3.1365 & 3.1208 & 3.1108 & 3.1041 & 3.0701   \\
$[6,3]$ &             & 3.8129 & 3.5003 & 3.3521 & 3.2725
        & 3.2255 & 3.1956 & 3.1756 & 3.1615 & 3.1078  \\
$[7,0]$ & 5.0307 & 4.0860 & 3.6934 & 3.4970 & 3.3852
        & 3.3155 & 3.2691 & 3.2369 & 3.2135 & 3.1048  \\
$[7,2]$ & 5.0319 & 4.0892 & 3.6992 & 3.5056 & 3.3963
        & 3.3287 & 3.2842 & 3.2534 & 3.2312 & 3.1366  \\
$[6,5]$ &             &             &             & 3.3598 & 3.2897
        & 3.2518 & 3.2299 & 3.2163 & 3.2075 & 3.1633   \\
$[7,4]$ &             &             & 3.7035 & 3.5159 & 3.4130
        & 3.3513 & 3.3119 & 3.2853 & 3.2666 & 3.1980   \\
\end{tabular}
%\end{ruledtabular}
\end{table*}

\begin{figure}
\includegraphics[width=8.5cm]{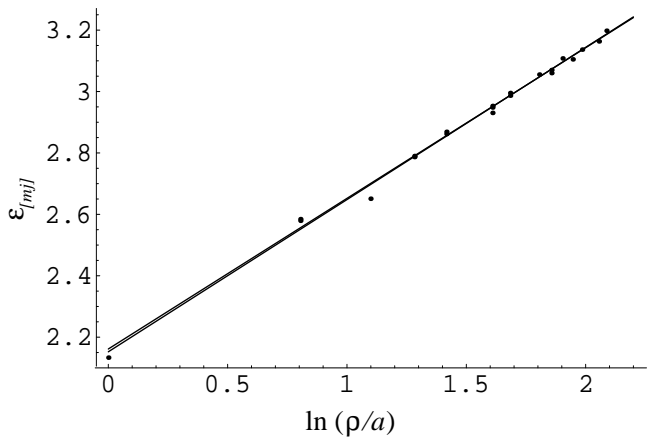}
\caption{\label{fig:dependencia}Energy, $\varepsilon_{[m,j]}$, of
Eq.~(\ref{eqn:energy-mj})
of a static pair of TSDs versus $\ln (\rho /a)$.
The continuous lines are linear series approximations:
(a) when all the points are retained 
[$\varepsilon_{[m,j]}=2.15235 +0.495788\ln (\rho /a)$];
(b)  when $(\rho /a) \leq 3$ are not included
[$\varepsilon_{[m,j]}=2.16149 +0.490593\ln  (\rho /a)$].}
\end{figure}

For moderate to long distances, our results indicate that the
Heisenberg energy of such a static excitation increases as
$\sim \ln \rho$, as predicted in Ref.~\onlinecite{PRB-91}.
Nevertheless, a tiny deviation from this behavior is observed for
small values of $\rho$.
This is because details of the lattice are more important for short
distances, as also is expected from the form of the denominator
in Eq.~(\ref{eqn:energy-mj}).
See, for instance, Fig.~\ref{fig:dependencia}.
Therefore, it is expected that the TSDs of a pair will try to remain
as close as possible. 
However, this is not enough to decide whether charge-wearing TSDs
organize themselves as bound pairs of two charge-wearing TSDs, with
bosonic character, or each charge-wearing TSD would bind to a
spin-wearing TSD, leading to a fermionic-character pair.

\section{Two-dimensional extended wave functions}
\label{sec:2D-wf}

The bosonic or the fermionic pairs are far from being
static.
The hopping terms of the Hamiltonian allow any charge-wearing TSD to
move while the exchange part mixes up all the VB configurations.
Therefore, the appropriate wave function must be a weighted
superposition of all possible static configurations, fulfilling
translational and point group symmetry conditions.
Thence, the wave functions for both bosonic and fermionic extended
pairs of TSDs should be invariant under the operations of the
factor group isomorphic to the $C_{4v}$ (4 mm) group,
as obtained by factorizing the full group into the translational
subgroup and the planar subgroup.
Thence, there can be conceivable extended wave functions
with symmetry $\mathcal{S}$,
\begin{equation}
{\mathcal{S}}= \left\{
\begin{array}{ll}
\mathcal{A}_1, & \text{totally symmetric},\: (x^2 +y^2),   \\
\mathcal{A}_2, & \text{antisymmetric under the four reflections}, \\
\mathcal{B}_1, & \text{antisymmetric under $C^\pm_4$,
$\sigma_{x \pm y}$},  \: (x^2 - y^2), \\
\mathcal{B}_2, & \text{antisymmetric under $C^\pm_4$,
$\sigma_x$, $\sigma_y$}, \: (xy).
\end{array} \right.
\end{equation}
Therefore, symmetry-adapted extended wave functions for both
the fermionic pairs and bosonic pairs can be written as 
\begin{equation}
\phi^{\mathcal{S}}_{[m,j]} (\mathbf{k}) \equiv
N_{[m,j]} \sum^{\in \mathcal{L}}_{\mathbf{R}}
\text{e}^{\text{i}\mathbf{k}\cdot \mathbf{R}}
\sum_{\mbox{\boldmath$\rho$}_{[m,j]}} | \mathbf{R},
\mathbf{R}+ \mbox{\boldmath$\rho$}_{[m,j]}\rangle
\chi^{\mathcal{S}}_{\mbox{\boldmath$\rho$}_{[m,j]}},
\label{eqn:wf}
\end{equation}
where $N_{[m,j]}$ is the normalization;
$\mbox{\boldmath$\rho$}_{[m,j]}$ is a vector obtained by any operation
of the point group acting on $[m,j]$, with $0\leq j<m$ and $m+j$=odd; 
$|\mathbf{R}, \mathbf{R}+ \mbox{\boldmath$\rho$}_{[m,j]}\rangle$ is
the state with a static pair of TSDs, let's say, a charge-wearing TSD
lying on site $\mathbf{R}$, and a spin-wearing TSD (a second
charge-wearing TSD) on $\mathbf{R}+ \mbox{\boldmath$\rho$}_{[m,j]}$
for the fermionic (bosonic) pairs;
$\chi^{\mathcal{S}}_{\mbox{\boldmath$\rho$}_{[m,j]}}$ is the
appropriate character of the irreducible representation $\mathcal{S}$,
with $\chi_{[m,j]}\equiv 1$.
Finally, $\mathcal{L}$ is the square lattice
($\mathcal{L}=\mathcal{A}$) for the fermionic (bosonic) pair.
Then, care must be taken with the allowed values for $\mathbf{k}$.
For instance, when dealing with the fermionic pairs, $\mathbf{k}$
belongs to the Brillouin zone of a square lattice with lattice
constant $a$.
On the other hand, for bosonic pairs $|k_x|, |k_y| \leq \pi /2a$,
because the summation is restricted to run on the sublattice
$\mathcal{A}$.

It can be readily seen that only the $\mathcal{A}_1$ and
$\mathcal{B}_1$ symmetries are allowed for $j$=0.
Since different symmetries do not mix, here we restrict ourselves to
$\mathcal{A}_1$ and $\mathcal{B}_1$ symmetries even when
$j\neq 0$.

\subsection{Energy of the fermionic pairs}

\begin{figure}
\includegraphics[width=8.5cm]{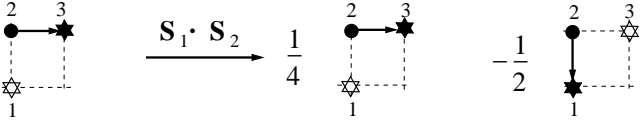}
\caption{The hopping of a spin-wearing TSD (white
star) by the action of $H_J$. }
\label{fig:SStransfers}
\end{figure}

The expectation values given by the wave functions of
Eq.~(\ref{eqn:wf}) are
\begin{eqnarray}
\zeta^{\mathcal{S}}_{[m,j]}(\mathbf{k})  &  \approx &
I -\frac{1}{2} t \left( \cos k_x a
+ \cos k_y a \right) \delta^{[1,0]}_{[m,j]} \nonumber \\
& + & \delta_{j,m-1}\left( 1+\delta^{[1,0]}_{[m,j]} \right)
 t'_1 \chi^{\mathcal{S}}_{[j,m]} \cos k_x a \cos k_y a
 \nonumber  \\
& + & J \left( \varepsilon_{[m,j]}+ \gamma_{[m,j]} \right) ,
\end{eqnarray}
where $I$ is the ionization potential (electron affinity) for
hole-doped (electron-doped) materials.
$t$ and $t'_i$ are nearest- and next-nearest-neighbor hopping
integrals, with $i=1$ when there is only one charge-wearing TSD in
the plaquette, and $i=2$ when there are two nearest-neighbor
charge-wearing TSDs in the plaquette;
$\gamma_{[m,j]}$ arises from the Heisenberg terms involving the
spin-wearing TSD (see fig.~\ref{fig:SStransfers}),
\begin{equation}
\gamma_{[m,j]} \approx \left\{
\begin{array}{ll}
\frac{3}{4} - \frac{1}{3} \chi^{\mathcal{S}}_{[0,1]},
 & m=1, \\
1-\frac{1}{8}\delta_{j,1}
-\frac{1}{4} \chi^{\mathcal{S}}_{[j,m]}\delta_{j,m-1},
  & \text{otherwise}\, .
\end{array}
\right.
\end{equation}

There are two families of nonzero off-diagonal matrix elements of the 
Hamiltonian.  When $m-m'= \pm 1$ and $j-j'=\pm 1$,
\begin{equation}
H^{[m',j']}_{[m,j]} = \lambda \left( t'_1 \cos k_x a \cos k_y a 
- \frac{1}{4}J- \frac{1}{8}J\delta_{j,m-1}\chi^{\mathcal{S}}_{[j,m]}
\right) .
\end{equation}
When $m'-m = \pm 2$ and $j'=j$ or $m'=m$ and $j'-j = \pm 2$,
\begin{equation}
H^{[m',j']}_{[m,j]} =\left\{ 
\begin{array}{ll}
-\sqrt{3}J/12, & [m',j']\text{ or } [m,j] = [1,0],\\
-\lambda J/8, & \text{otherwise}, 
\end{array}
\right.
\end{equation}
with $\lambda = \sqrt{2}$ when either $j$ or $j'$ is zero (but not both),
and $\lambda = 1$ otherwise.

The zero-order lowest-lying fermionic pairs are the $[1,0]$.
Close to $\Gamma$ the energy of these fermionic pairs is
\begin{equation}
\label{eqn:xi-aprox}
\zeta^{\mathcal{S}}_{[1,0]}(\mathbf{k})  \approx
\zeta^{\mathcal{S}}_\Gamma 
+ \frac{\hbar^2 k^2_\parallel }{2m^{\mathcal{S}}_\parallel }
\end{equation}
with
\begin{eqnarray}
\label{eqn:xi-Gamma}
\zeta^{\mathcal{S}}_\Gamma & \equiv  &I - t
+2 t'_1 \, \chi^{\mathcal{S}}_{[0,1]}
+J\left( \varepsilon_{[1,0]}+ \frac{3}{4} - \frac{1}{3} 
\chi^{\mathcal{S}}_{[0,1]} \right) \nonumber \\[2mm]
\frac{\hbar^2 }{2m^{\mathcal{S}}_\parallel } &\equiv   &
\left( \frac{1}{4} t - t'_1 \chi^{\mathcal{S}}_{[0,1]}
\right) a^2
\end{eqnarray} 

Thence, for $t'_1>J/6$, the zero-order lowest-lying band has $x^2-y^2$
symmetry.
This is the case for the La$_{2-x}$Sr$_x$CuO$_4$
(Ref.~\onlinecite{Calzado}) (LSCO).
For the monolayered HTSC of Table 1 in Ref.~\onlinecite{CPL} we know
that $J/6 \sim 0.019\text{---}0.030$~eV.
On the other hand, strong differences on hopping integrals among the
different HTSC are not expected, as suggested by the small variations
observed on the nearest-neighbor hopping integral, $t$.
Therefore, we expect that the zero-order lowest-lying band will show 
$\mathcal{B}_1$ symmetry for all of these HTSC.

We are now concerned whether admixing wave functions with different
$[m,j]$ to the $\phi^{\mathcal{S}}_{[1,0]}(\mathbf{k})$ would be
relevant or even if the ordering of the lowest-lying $\mathcal{A}_1$
and $\mathcal{B}_1$ bands could be reversed.
We have obtained the corrections to
$\zeta^{\mathcal{A}_1}_{[1,0]}(\mathbf{k})$ and
$\zeta^{\mathcal{B}_1}_{[1,0]}(\mathbf{k})$ by diagonalizing
the matrix of the Hamiltonian in the basis of the two, three
and four lowest-lying wave functions with the appropriate
$\mathcal{A}$ or $\mathcal{B}$ symmetries.
Thence, making use of the parameters for the
LSCO,\cite{Calzado} when the number of fermionic pairs per Cu is
$p= 0.05 \text{---}0.07$ we obtain corrections to the zero-order Fermi
energy of
$\Delta^{(2)}=-16.98$ to $-14.56$ meV,
$\Delta^{(3)}=-1.45$ to $-1.33$ meV, and
$\Delta^{(4)}=-0.26$ to $-0.20$ meV for the $\mathcal{B}_1$ band,
while for the $\mathcal{A}_1$ the corrections are
$\Delta^{(2)}=-7.21$ to $-5.1$ meV,
$\Delta^{(3)}=-3.15$ to $-2.7$ meV, and
$\Delta^{(4)}=-0.15\text{ to }-0.1$ meV.
For the monolayered HTSC of Table 1 in Ref.~\onlinecite{CPL}, assuming
$t'_1 \sim 0.2 t$, these corrections are slightly decreasing with
$J/t$.
We observe that, up to meV, the correction to the $\mathcal{A}_1$ energy
is smaller than the correction to the $\mathcal{B}_1$.
Therefore, it is expected that the lowest-lying band still has
$x^2-y^2$ symmetry.
Furthermore, the band with symmetry $\mathcal{A}_1$ would
not start filling until a critical doping of
$p \approx 0.20\text{---}0.22$ holes per CuO$_2$ unit,
provided that all the charge-wearing TSDs organize as fermionic pairs.
At this doping, the corrections $\Delta^{(n)}$ to the energies are
still smaller than those referred above.
Since this doping is out of the range of our interest, we restrict
ourselves to consider only the band with $x^2-y^2$ symmetry.
Also, since the error in the parameters of the Hamiltonian are of the
order of meV, we neglect corrections to the energy smaller than 1 meV.
Therefore, we consider only the $[1,0]$, $[2,1]$, and $[3,0]$ wave
functions to describe the lowest-lying band of the fermionic pairs.

\subsection{Energy of the bosonic pairs}

When dealing with $t$-$J$-like model Hamiltonians, the electrostatic
repulsion is generally neglected, although with some
exceptions.\cite{Kivelson90,Dagotto92}
Since the \emph{screened} electrostatic repulsion between the
charge-wearing TSDs in a pair, $V_{[m,j]}$, may be relevant,
here it is included in the diagonal terms of the Hamiltonian,
\begin{equation}
\langle H\rangle^{\mathcal{S}}_{[m,j]}
= J \varepsilon_{[m,j]}+ 2I  + V_{[m,j]} +
\langle H_t +H_{t'} \rangle ^{\mathcal{S}}_{[m,j]}.
\end{equation}
It is not difficult to obtain
$\langle H_t  \rangle ^{\mathcal{S}}_{[m,j]}= 0$,
and
\begin{equation}
\langle  H_{t'}\rangle ^{\mathcal{S}}_{[m,j]} =  \kappa_{[m,j]}
\chi^{\mathcal{S}}_{[j,m]}
\left( 1+\cos k_x a \cos k_y a \right) ,
\end{equation}
with
\begin{equation}
\kappa_{[m,j]} = \left\{
\begin{array}{ll}
2t'_2,  & [m,j]=[1,0], \\
t'_1,  &  m-j=1,  \: m>1, \\
0,  & \text{otherwise}.
\end{array}
\right.
\end{equation}

The electrostatic repulsion within a bosonic pair is expected to be
screened by the gas of the fermionic pairs.
However, the fermionic pairs have been considered so far as
moving in a two-dimensional square lattice.
It is generally accepted that the $c$ axis effect is simply to tune
the electronic structure of the CuO$_2$ planes.
Nevertheless, screening is a three-dimensional effect that could be
taken into account by an interlayer hopping integral $t_\perp$.
Considering a nonzero $t_\perp$ would imply a correction to the
energy $\approx t_\perp k^2_\perp c^2$, where $c$ is the lattice
constant perpendicular to the $ab$ layers.
Since $t_\perp$ is rather small,\cite{Zha} it can be neglected for
the energy-balance considerations, but it is essential for screening
purposes.
Then, for the electrostatic repulsion within a bosonic pair we take as
a first approximation the Yukawa potential\cite{Ashcroft,Madelung} as
the dominant term,
\begin{equation}
V_\rho \approx \frac{q^2}{\rho}
\exp \left\{ -\left( 4 \pi e^2 g_F \right)^{1/2}
 \rho\right\} \, ,
\end{equation}
where $g_F$ is the density of states at the Fermi
level of the fermionic pairs per unit of volume of the solid,
\begin{equation}
g_F \approx
\left( \frac{3m_\perp m^2_\parallel  \nu p}{\pi^4 \hbar^6 a^2 c}
\right)^{1/3}\, ,
\end{equation}
$p$ being the number of fermionic pairs per site, and
$\nu$ is the number of square-lattice layers cutting a unit cell.
Therefore, close to $\Gamma$, the diagonal terms of the Hamiltonian
are
\begin{eqnarray}
\langle H\rangle^{\mathcal{S}}_{[m,j]}
& \approx & J \varepsilon_{[m,j]}+ 2I +  \frac{e^2}{\rho}
\exp \left\{ - \beta\rho  (\nu p)^{1/6}  \right\} \nonumber \\
&+&\kappa_{[m,j]} \chi^{\mathcal{S}}_{[j,m]}
\left( 2-\frac{1}{2}a^2k^2\right) \, ,
\label{eqn:zeta-Gamma}
\end{eqnarray}
with
\begin{equation}
\beta \approx  \frac{2  e}{a} \left(
\frac{6}{\pi t_\perp (t+4 t'_1)^2 c^3 } \right) ^{1/6}\, .
\end{equation}
The non-zero off-diagonal elements of the Hamiltonian can also be
readily obtained,
\begin{equation}
H^{[m',j']}_{[m,j]} =
\lambda t'_1 (1+ \cos k_x a \cos k_y a ); \: |m-m'|, |j-j'|=1,
\end{equation}
with $\lambda = \sqrt{2}$ when either $j$ or $j'$ is zero, and
$\lambda = 1$ otherwise.

Since the nonzero off-diagonal elements of the Hamiltonian are
important as compared to the differences among the diagonal elements,
the energy of the bosonic pairs must be obtained by
diagonalizing the matrix of the Hamiltonian.

Since the screened electrostatic repulsion decays faster than
exponentially, while $\varepsilon_{[m,j]}$ increases
logarithmically, there must be a minimum in the energy and
confinement is expected to occur.
Furthermore, from Eq.~(\ref{eqn:zeta-Gamma}) and assuming that $t'_1$,
$t'_2> 0$,\cite{Calzado} it is expected that the bosonic pairs will
also show $x^2- y^2$ symmetry as it is generally
accepted.\cite{Dagotto94}
At this point, it is worth noting that if the next-nearest-neighbor
hopping is neglected and the Hamiltonian is reduced to the $t$-$J$
model, it turns out that the $\mathcal{A}_1$ and $\mathcal{B}_1$
symmetries would be degenerate.

\subsection{Two-fluids equilibrium condition}

At $T=0$, the question now is whether this lowest-lying bosonic pairs
would have lower energy than two fermionic pairs in its Fermi level,
so the bosonic pairs would be favored.
At low doping level, it is expected that the fermionic pairs will be
favored.
Nevertheless, the Fermi level, $k_F \approx \sqrt{2\pi p}/a$,
increases linearly with $p$, while the electrostatic repulsion among
the two charge-wearing TSDs in the bosonic pair is exponentially
reduced with $p^{1/6}$, so its ground-state energy is lowered.
Therefore, we wonder whether there exist a critical value $p_c$ such
that the ground-state of the bosonic pairs, as measured with respect
twice the energy of a fermionic pair at its Fermi level,
$\Delta(p)$, is zero.

To explore such a possibility, we have diagonalized the matrix of the
Hamiltonian for $\mathbf{k}=0$, and increasing values of $p$.
To reach corrections to the ground-state energy within the order of
meV, we have considered up to a $12\times 12$ matrix involving all the
states which would contribute to perturbation theory truncated to
tenth order.   
For a certain regime of parameters, at low enough doping level, the
fermionic pairs are favorable.
As $p$ increases there exist a critical value of doping, $p_c$, such
that $\Delta(p_c) = 0$ for the lowest-lying bosonic pairs.

Doping above $p_c$ yields bosonic pairs.
In this case, the  bosonic pairs are expected also to contribute to
the screening and the electrostatic repulsion could become
negligible.
If so, there would be a cascade process of pairing among the fermionic
pairs until a new equilibrium between the two fluids is reached at
$p_f<p_c$.
Thence, we expect the number of bosonic pairs at $p>p_f$ to be
$p_b\approx (p-p_f)/2$.
There is a lower limit of $p_f$ such that $\Delta_0(p_f) = 0$, as
obtained when the electrostatic repulsion is completely neglected.

\begin{figure}
\includegraphics[width=8.5cm]{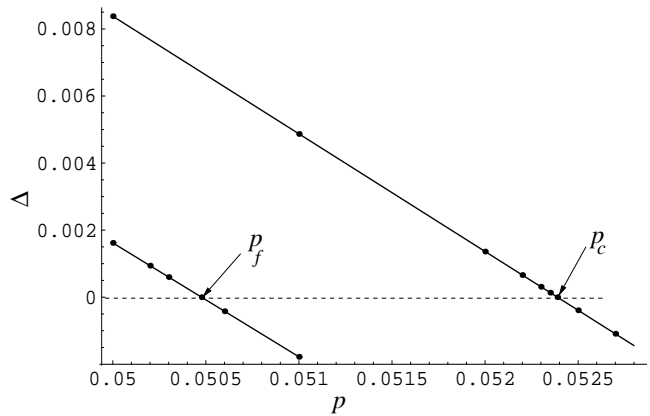}
\caption{\label{fig:Dp_c_f} The ground-state energy of the bosonic
pairs, as measured respect twice the energy of a fermionic pair at its
Fermi level, $\Delta(p)$.  Top, $\Delta (p)$, with screened
electrostatic repulsion. Bottom, $\Delta_0(p)$, without
electrostatic repulsion.}
\end{figure}

For the sake of estimating the order of magnitude of $p_c$
and $p_f$ for a generic HTSC, let us make use of the parameters
appropriate for LSCO.
We take $a\approx 3.8$~\AA\ and $c/a \approx 3.47$
from Ref.~\onlinecite{Dagotto94}.
For the Hamiltonian parameters, we take $J=0.144$~eV and
$t=0.549$~eV from Ref.~\onlinecite{CPL}, and $t'_2=0.130$~eV and
$t'_1=0.112$~eV from Ref.~\onlinecite{Calzado}.
All of these parameters were obtained from high-level
\emph{ab initio} calculations with the geometry as the only
external input, being the errors within the meV.
Since there is no high-level
\emph{ab initio} calculation for the interlayer
hopping integral, we use the low-doping $t_\perp \approx 0.7$~meV
obtained from experimental results by Zha, Cooper, and Pines.\cite{Zha}
Within these parameters regime we get $\beta \approx 7.8/a$.
Computing $\Delta(p)$ for increasing values of $p$ we find that
$\Delta(p)$ is changing its sign at $p_c \approx 0.0524$
(see, for instance, Fig.~\ref{fig:Dp_c_f}).
At $p_c$, the mean distance between the two holes of the pair is
$\langle \rho_c \rangle \approx 9.08$~\AA\ with a standard deviation
$\sigma_c \approx 5.08$~\AA .
Identifying in a rather loose way the spatial extent of the pair
wave function ($\sim \rho_c +\sigma_c$) with the coherence length
$\xi$, we obtain $\xi \approx 14.16$~\AA , in good agreement with
the in-plane value ($\xi \sim 14$-$15$~\AA )\cite{Barnes,Dagotto94}
suggested from experimental findings.

On the other hand, when the electrostatic repulsion is neglected,
$\Delta_0(p)$ changes its sign at $p_f \approx 0.0505$, and the
mean distance between the two holes is
$\langle \rho_f \rangle \approx 8.78$~\AA\ with a standard deviation
$\sigma_f \approx 4.99$~\AA .
Therefore, the estimated coherence length is $\xi \approx 13.77$~\AA .
Again, it is worth noting that if the next-nearest-neighbor hopping
is neglected and the Hamiltonian is reduced to the $t$-$J$ model,
the value of the critical doping ($p_c \approx 0.28$) is out of the
range where the superconductivity is observed.
In addition, at so high doping the validity of such model
Hamiltonians could be questioned.

\begin{figure}
\includegraphics[width=8.2cm]{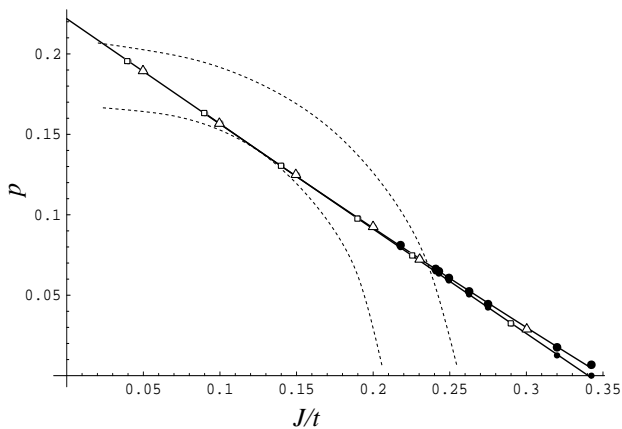}
\caption{$p_c$ for the monolayered HTSC (\textbullet ) and for LSCO
with different $J/t$ ($\triangle$);  $p_f$  for the monolayered HTSC
({\large $\bullet$}) and for LSCO with different $J/t$ ($\Box$).
The continuous lines are power series approximations to $p_c$ and
$p_f$ sets.  The dotted lines are hypothetical $J/t$ vs $p$ curves.}
\label{fig:p-J-t}
\end{figure}

As far as we know, for other monolayered cuprate superconductors only
the $t$ and $J$ parameters have been obtained from high-level
\emph{ab initio} calculations.
Nevertheless, for the purpose of estimating how $p_c$ and $p_f$
vary with $J/t$, let us assume that the ratios $t'_1/t$ and $t'_2/t$,
as well $a \beta$ do not change very much among them, and use the values
appropriate for LSCO.
If so, we find that $p_f$ and $p_c$ increase as $J/t$ decreases.
See, for instance, Fig.~\ref{fig:p-J-t}, where we also include the
$p_f$ and $p_c$ values for LSCO as a function of $J/t$.
This result suggests that for low $J/t$ the onset of
superconductivity would be located at a too high level of doping
such that it could be beyond the validity of the present
approximation.
Therefore, the superconductivity could be suppressed.

Furthermore, it is worth noting here that the parameters that
characterize a superconductor are taken as independent of doping.
Nevertheless, $J$ as well as the hopping integral do depend locally on
doping, as suggested by the high-level
ºemph{ab initio} calculations of Calzado
and Malrieu.\cite{Calzado}
For instance, their calculations suggest that $J$ decreases, while $t$
increases, with doping.
Therefore, it is expected that $J/t$ decreases with doping, probably
not linearly (see, for instance, the dotted lines of
Fig.~\ref{fig:p-J-t}).
Consequently, $p_b$ would decrease, eventually down to zero at a
critical doping $p'_c$ such that the $J/t$ as a function of the doping
crosses again the $p_f$ function.
Thence, the superconductivity would be suppressed in the overdoped
regime for a doping $p>p'_c$.
Therefore, a better knowledge of the parameters is essential to fully
understand the phenomenon of HTSC.

\section{Summary and Conclusions}
\label{sec:conclusions}

We have shown that the Heisenberg energy associated with a pair of TSD
in a spin=1/2 square lattice increases logarithmically with distance.
Therefore, a charge-wearing TSD (either a hole or a doubly occupied
site) in a spin=1/2 square lattice binds to another TSD, either to a
spin-wearing TSD or to another charge-wearing TSD.

We have constructed symmetry-adapted extended wave functions for
both a fermionic pair of a charge-wearing TSD and a spin-wearing
TSD, and a bosonic pair of two charge-wearing TSD.
The energy associated with such fermionic and bosonic pairs has been
obtained.

For the lowest-lying fermionic band and the lowest-lying bosonic pairs
the symmetry turns to be $x^2-y^2$ when $t'_ 1> J/6$ and 
$t'_1, t'_2> 0$, respectively.
Since these conditions are fulfilled for monolayered HTSC, we obtain
that the symmetry of the bosonic pairs is $x^2-y^2$ for these
materials, as it is generally accepted.\cite{Dagotto94}

For the LSCO compound, we find a critical doping for bosonic pairing
$p_c\approx 0.0524$ and $p_f\approx 0.0505$ (when the electrostatic
repulsion is completely neglected). 
This finding could be related to the onset of High $T_\text{c}$
superconductivity, the superconducting state being a Bose condensate.
This is also compatible with the existence of pairs above
$T_\text{c}$, a forerunner of the pseudogap physics of the cuprates.
At the critical doping, we find a mean distance between the two holes
of the pair $\langle \rho_c \rangle \approx 9.08$~\AA\ %
($\langle \rho_f \rangle \approx 8.78$~\AA ), and an estimated
coherence length $\xi \approx 14.16$~\AA\ ($\xi \approx 13.77$~\AA ).
These features are in good agreement with the
experimental result of $p_c \approx 0.05$ (Ref.~\onlinecite{Yamada})
and $\xi \sim 14$-$15$~\AA .\cite{Barnes,Dagotto94}

For the monolayered cuprate superconductors of Table 1 in
Ref.~\onlinecite{CPL}, we have obtained $p_f$ and $p_c$ as a function
of $a$ and $J/t$, while keeping fixed $a \beta \approx 7.8$ and
$t'_i/t$.
See, for instance, Fig.~\ref{fig:p-J-t}.
It can be observed that $p_f$ and $p_c$ increase as $J/t$ is lowered.

Extensions of the present work towards $T\neq 0$, and to other
possible charge-wearing TSDs self-organization are in progress.
For instance, it is of interest to explore
the parameter regime for bosonic pairs with $xy$ symmetry,
phase separation, and stripe formation.

\begin{acknowledgments}

This research was supported by the Spanish
DGI (Project No.~PFM2002-02629).
the author thanks Professor D.J. Klein, Professor A. Labarta, Professor
F. Illas and Dr.\ I. de P. R. Moreira for valuable suggestions.

\end{acknowledgments}


\begin{thebibliography}{99}
\bibitem{Bednorz}  J. G. Bednorz and K. A. M\"uller, Z. Phys.\ B:
Condens. Matter \textbf{64}, 188 (1986).
\bibitem{Barnes}  T. Barnes, Int.\ J. Mod.\ Phys.\ C \textbf{2}, 659
(1991).
\bibitem{Dagotto94}  E. Dagotto, Rev.\ Mod.\ Phys.\ \textbf{66}, 763
(1994).
\bibitem{Orenstein}  J. Orenstein and A. J. Mills, Science
\textbf{288}, 468 (2000).
\bibitem{NCYeh}  Nai-Chang Yeh, cond-mat/0210656.
\bibitem{Norman}  M. R. Norman and C. P\'epin,
Rep.\ Prog.\ Phys.\ \textbf{66}, 1547 (2003). 
\bibitem{Perdew} J. P. Perdew, in \emph{Electronic Structure of
Solids}, edited by P. Ziesche and H. Eschrig (Akademic Verlag,
Berlin, 1991).
\bibitem{Pickett}  W. E. Pickett, Rev.\ Mod. Phys.\ \textbf{61}, 433
(1989); \textbf{61}, 749E (1989).
\bibitem{Moreira} I. de P.R. Moreira and R. Dovesi,
Int.\ J. Quantum Chem.\ \textbf{99}, 805 (2004).
\bibitem{Godby} R.W. Godby, M. Schluter, and L.J. Sham,
Phys.\ Rev. Lett.\ \textbf{56}, 2415 (1986).
\bibitem{Hybertsen86} M. S. Hybertsen and S.G. Louie,
Phys.\ Rev.\ B \textbf{34}, 5390 (1986).
\bibitem{S-Gunnarsson}  A. Svane and O. Gunnarsson,
Phys.\ Rev.\ Lett.\ \textbf{65}, 1148 (1990).
\bibitem{Svane} A. Svane, Phys.\ Rev.\ Lett.\ \textbf{68}, 1900
(1992).
\bibitem{Szotek}  Z. Szotek, W. M. Temmerman, and H. Winter,
Phys.\ Rev.\ B \textbf{47}, 4029 (1993).
\bibitem{Czyzyk}  M. T. Czyzyk and G. A.Sawatzky,
Phys.\ Rev.\ B \textbf{49}, 14211 (1994). 
\bibitem{Anisimov}  V. I. Anisimov, M. A. Korotin,
J. Zaanen, and O. K. Andersen,
Phys.\ Rev. Lett.\ \textbf{68}, 345 (1992). 
\bibitem{Wei}  Pan Wei and Zheng Qing Qi,
Phys.\ Rev.\ B \textbf{49}, 12159 (1994).
\bibitem{Anderson87} P.W. Anderson, Science \textbf{235}, 1196 (1987).
\bibitem{Emery87}  V. J. Emery, Phys.\ Rev. Lett.\ \textbf{58}, 2794
(1987).
\bibitem{Emery88}  V. J. Emery and G. Reiter, Phys.\ Rev.\ B
\textbf{38}, 4547 (1988).
\bibitem{Zhang}  F. C. Zhang and T. M.Rice, Phys.\ Rev.\ B \textbf{37},
3759 (1988).
\bibitem{CPL}  I. de P. R. Moreira, D. Mu\~noz, F. Illas,
C. de Graaf, and M. A. Garcia-Bach, Chem.\ Phys.\ Lett.\ \textbf{345},
183 (2001).
\bibitem{Calzado}  C.J. Calzado and J.-P. Malrieu,
Phys.\ Rev.\ B \textbf{63}, 214520 (2001).
\bibitem{Munoz}  D. Mu\~noz, I. de P. R. Moreira, and F. Illas,
Phys.\ Rev.\ B \textbf{65}, 224521 (2002).
\bibitem{Lieb-Mattis}  E. H. Lieb and D. C. Mattis, J.
Math.\ Phys.\ \textbf{3}, 749 (1962).
\bibitem{KivelsonRS}  S. A. Kivelson, D. S. Rokhsar, and J. P. Sethna,
Phys.\ Rev.\ B \textbf{35}, 8865 (1987).
\bibitem{Kivelson}  S. Kivelson, Phys.\ Rev.\ B \textbf{36}, 7237
(1987).
\bibitem{PRB-91}  D. J. Klein, T. G. Schmalz, M. A. Garcia-Bach,
R. Valent\'{\i}, and T. P. \v{Z}ivkovi\'c, Phys.\ Rev.\ B \textbf{43},
719 (1991).
\bibitem{KleinZV}  D. J. Klein, T. P. \u{Z}ivkovi\'c, and R. Valent\'i,
Phys.\ Rev.\ B \textbf{43}, 723 (1991).
\bibitem{Hansson}  T. H. Hansson, V. Oganesyan, and S. L. Sondhi,
Ann.\ Phys. (N.Y.) \textbf{313}, 497 (2004). 
\bibitem{EPJ} M. A. Garcia-Bach, Eur.\ Phys.\ J. B \textbf{14}, 439
(2000).
\bibitem{capitolVB}  M. A. Garcia-Bach, ``Many-body VB
ans\"atze.  From polymers and ladder materials to the square lattice,''
\emph{Valence Bond Theory}, (Elsevier, New York, 2002),
p.~729-768.
\bibitem{Hybertsen90}  M. S. Hybertsen, E. B. Stechel, M. Schluter,
and D. R. Jennison,   Phys.\ Rev.\ B \textbf{41}, 11068 (1990).
\bibitem{Ashcroft}  N. W. Ashcroft and N. D. Mermin,
\textit{Solid State Physics} (Holt-Saunders, Japan, 1981),
pp.\ 337-342.
\bibitem{Madelung} O. Madelung, \textit{Introduction to
Solid-State Theory}, Springer Series in Solid State Sciences,
Vol.\ \textbf{2}, (Springer-Verlag, Berlin, 1978), pp.\ 114-117.
\bibitem{Seitz}  W. A. Seitz, D. J. Klein, T. G. Schmalz, and M. A.
Garcia-Bach, Chem.\ Phys.\ Lett.\ \textbf{115}, 139 (1985);
\textbf{118}, 110E (1985).
\bibitem{Klein-86}  D. J. Klein, G. E. Hite, and  T. G. Schmalz,
J. Comput.\ Chem.\ \textbf{7}, 443 (1986).
\bibitem{Zivkovic}  T. P. \v{Z}ivkovi\'c, B. L. Sandleback, T. G.
Schmalz, and D. J. Klein, Phys.\ Rev.\ B \textbf{41}, 2249 (1990).
\bibitem{Liang}  S. Liang, B. Doucot, and P. W. Anderson,
Phys.\ Rev.\ Lett.\ \textbf{61}, 365 (1988).
\bibitem{Kivelson90}  S. A. Kivelson, V. J. Emery, and H. Q. Lin,
Phys.\ Rev.\ B \textbf{42}, 6523 (1990). 
\bibitem{Dagotto92} E. Dagotto and J. Riera, Phys.\ Rev.\ B
\textbf{46}, 12084 (1992).
\bibitem{Zha}  Y. Zha, S.L. Cooper, and D. Pines, Phys.\ Rev.\ B
\textbf{53}, 8253 (1996). 
\bibitem{Yamada}  K. Yamada \emph{et al.},  Phys.\ Rev.\ B
\textbf{57}, 6165 (1998).


\end{thebibliography}
\end{document}